# Chapter 17
# The Evolution of Intelligence

How did the human species evolve the capacity not just to communicate complex ideas to one another but to hold such conversations from across the globe, using remote devices constructed from substances that do not exist in the natural world, the raw materials for which may have been hauled up from the bowels of the earth? How did we come to be so intelligent? Research at the interface of psychology, biology, anthropology, archaeology, and cognitive science is culminating in an increasingly sophisticated understanding of how human intelligence evolved. Studies of the brains of living humans and great apes and the intellectual abilities they support are enabling us to assess what is unique about human intelligence and what we share with our primate relatives. Examining the habitats and skeletons of our ancestors gives cues as to environmental, social, and anatomical factors that both constrain and enable the evolution of human intelligence. Relics of the past also have much to tell us about the thoughts, beliefs, and abilities of the individuals who invented and used them.

The chapter starts with an introduction to some key issues in the evolution of intelligence. We then consider what is unique about human intelligence compared to our closest living biological relatives, the great apes – chimpanzees, bonobos, gorillas, and orangutans. The process by which the human intelligence came about is the next topic. Finally, we address the question of *why* human intelligence evolved – did it evolve purely due to biological forces, that is, does intelligence merely help us solve survival problems and attract mates, or are nonbiological factors such as culture involved?

## Key Issues

We begin by laying out some of the fundamental issues that arise in considerations of the evolution of human intelligence. First, we address some issues of definition. Second, we comment on challenges to the accurate assessment of intelligence, particularly when comparing intelligence across different species. A third, related issue is the question of the extent to which there are special qualities of intelligence that only humans attain.

### Assessing Intelligence and Its Evolution

Many methods are used to assess intelligence and its evolution. These include (1) *behavioral measures*, which may involve naturalistic observation or analyzing responses in laboratory experiments, (2); *artifactual measures,* which involve analysis of tools, art, and so forth,; and (3) *anatomical/neurological measures*, which involve studies of the brain and cranium. Ideally, all three would converge upon a unified picture of how intelligence evolved. However, this is not always the case, and indeed, the assessment of intelligence is fraught with challenges.
An obvious one is that we cannot perform behavioral or neurological studies of our ancestors, so we are forced to rely on bones and artifacts. Moreover, the further back in time one looks, the more fragmentary the archaeological record becomes. To explore the ancestral roots of our intelligence, we therefore also partly rely on studying the intelligence and brains of the great apes, our closest biological relatives. We share a common ancestor with great apes as recently as 4–6 million years ago (mya): No living species are more closely related. Other species such as dolphins and crows share some complex intellectual abilities with great apes and humans, but their abilities probably evolved independently and operate differently. Dolphins' and crows' brains differ strikingly from ours, for instance, whereas great ape brains are exceptionally similar to ours Emery & Clayton, 2004; Hof, Chanis & Marino, 2005; MacLeod, 2004).

What the great apes offer to the study of the evolution of human intelligence is the best living model of the intelligence that existed in our common great ape ancestors before our unique evolutionary lineage, the hominins, diverged. Modern human intelligence evolved from earlier forms of intelligence in response to selective pressures generated by ancestral living conditions. Understanding its evolution therefore entails looking into the past for the changes that occurred within the hominins – but also for earlier intellectual traits upon which the hominins built and the changes that led to the their divergence from ancestral great apes. If we can identify complex behaviors that great apes share with humans but not with other nonhuman primates, then these behaviors and the intellectual qualities they imply may have been shared by our common ancestors.

To use great apes to contribute to understanding the evolution of human intelligence, especially inferring what intellectual capacities evolved uniquely in the hominins, we need to assess their intellectual ceiling, that is, their top adult-level capabilities near the human boundary. The intelligence of great apes is highly malleable and dependent on the developmental and learning history of the individual (Matsuzawa, Tomonaga & Tanaka, 2006; Parker & McKinney 1999; de Waal, 2001), as it is in humans. Conclusions about great ape cognition and comparisons with human cognition must therefore be made with care. In part because this care has not always been taken, the literature on how human intelligence evolved does not present as straightforward a picture as one might hope. Nevertheless, an integrated account is starting to emerge.

## What Distinguishes Human From Nonhuman Intelligence?

Many have attempted to specify what marks the intellectual divide between humans and other species. Some follow Aristotle's proposal that it is reason (French, 1994), or symbolic thinking. Symbols are arbitrary signs with conventional meanings that are used to represent (stand for) other things or relationships between them, and that generally have conventionally accepted meanings. Another suggestion is that human intelligence is distinguished by the ability to develop complex, abstract, internally coherent systems of symbol use (Deacon, 1997). Others propose that it is creativity, such as is required to invent tools, or abilities associated with creativity, such as or cognitive fluidity (combining concepts or ideas, or adapting them to new contexts), or the ability to generate and understand analogies (Fauconnier & Turner, 2002; Mithen, 1996). Still other proposals single out key abilities for dealing with the social world, such as demonstration teaching, imitative learning, cooperative problem solving, or communicating about the past and future. A related proposal is that the divide owes to the onset of what Premack and Woodruff (1978) refer to as *theory of mind*—the capacity to reason about mental states of others (Mithen, 1998).

The more we learn about nonhuman intelligence, however, the more we find that abilities previously thought to be uniquely human are not. For example, it was thought until the 1960s that humans alone make tools. But then Jane Goodall (1963) found wild chimpanzees making them. Later, several other species were found making tools too (Beck, 1980). Thus, ideas about what marks the boundary between human and nonhuman intelligence have undergone repeated revision.

Although a large gulf separates human abilities from those of other species, it is not as easy as we hoped to pinpoint in a word or two what distinguishes humans. That does not mean that a more complex explanation is not forthcoming. For example, it may be that it is not creativity per se that distinguishes human intelligence, but the proclivity to take existing ideas and adapt them to new contexts or to one's own unique circumstances – that is, to put one's own spin on them, such that they become increasingly complex. The question of what separates human intelligence from that of other species is a recurring theme that will be fleshed out in the pages that follow.

## Intelligence in Our Closest Relatives: The Great Apes

We now possess a rich body of data on great ape intelligence (Byrne, 1995; Gómez, 2004; Matsuzawa *et al.*, 2006; Parker & Gibson, 1990; Povinelli, 2000; Rumbaugh & Washburn, 2003; Russon, Bard & Parker, 1996; Tomasello & Call, 1997; de Waal, 2001). This section summarizes the current picture of great ape intelligence, focusing on qualities once thought to be uniquely human. While some monkeys have shown similar achievements, great apes consistently achieve higher levels (Parker & McKinney, 1999).

Great apes have shown many social cognitive abilities thought uniquely human. They show imitative learning and demonstration teaching powerful enough to sustain simple cultures (Boesch, 1991; Byrne & Russon, 1998; Parker 1996; van Schaik et al., 2003; Whiten et al., 1999). Some have solved problems cooperatively (Boesch & Boesch-Achermann, 2000; Hirata & Fuwa, 2007) and show some understanding of others' mental states (e.g., knowledge, competence) (Parker & McKinney, 1999). Captives have acquired basic sign language, including learning and inventing arbitrary conventional signs and simple grammar (Blake, 2004). Some great ape gestures qualify as symbolic by standards used in early language studies, including tree-drumming, holding thumb and finger together and blowing through them to represent a balloon, and making twisting motions toward containers they wanted opened (Blake, 2004).

Great apes can understand simple analogies and engage in analogical reasoning (Thompson & Oden, 2000). They are considered to achieve basic symbolic abilities in several problem domains; they can do simple arithmetic and master simple language, for example (Parker & McKinney, 1999; Thompson & Oden, 2000).

A certain degree of creativity may be normal in great apes (and other nonhuman species; Reader & Laland 2003). Their creativity includes smearing leaf pulp foam on their body (perhaps as an analgesic), inventing new tools (e.g., branch hook tools, termite fishing brush tools), primitive swimming, and fishing (Russon et al., 2009; Sanz & Morgan, 2004). They have invented gestures and signs such as hand shaking and tree drumming (Boesch, 1996; Goodall, 1986). Some have mimed inventively; examples are making hitting actions toward nuts they want cracked, blowing between thumb and forefinger to represent a balloon, and making twisting motions at containers they want opened (Miles et al., 1996; Russon, 2002; Savage-Rumbaugh et al., 1986).

One approach to assessing great ape intelligence is measuring their performances against children's on the same cognitive task. Chimpanzees can use scale models, for instance, which children first master in their third year (Kuhlmeier, Boysen & Mukobi, 1999). Chimpanzees and orangutans have solved *reverse contingency* tasks, which allow a subject to choose one of two sets of items (e.g., different amounts of candies) but then give the subject the set *not* chosen (Boysen et al., 1996; Shumaker et al., 2001). Chimpanzees who understood number symbols solved this task (chose the smaller amount to receive the larger) when amounts were shown by symbols, but failed with real foods. Children first solve this task between three and three and a half years of age and three year olds show limitations like the chimpanzees' (Carlson, Davis, & Leach, 2005). Thus some great apes show certain symbolic logical abilities comparable to those of three and a half year old children. To date, great apes have not shown evidence of the symbol systems that Deacon (1997) proposes to distinguish human intelligence.

## Summary and Implications of Great Ape Research for Human Intelligence

There is now fairly strong agreement that great apes share a grade of intelligence of intermediate complexity that goes beyond that of other nonhuman primates and includes abilities previously thought uniquely human (Byrne, 1995; Gómez, 2004; Langer, 1996; Matsuzawa, 2001b; Parker & McKinney, 1999, Russon, 2004). A minority of primatologists view great ape intelligence as not significantly different from that of other nonhuman primates, on the one hand, or as more powerful but not reaching the currently defined human boundary, on the other (e.g., Povinelli, 2000; Suddendorf & Whiten, 2002; Tomasello & Call, 1997). Disagreement is due partly to emphasizing weak performances, interpreting monkey evidence too generously, neglecting great apes' most complex achievements, or incorrectly discounting them as artificially boosted by human enculturation. All-in-all, however, the evidence remains consistent with Premack's (1988) rule of thumb: Under normal circumstances great apes can reach levels of intelligence of 3.5-year-old children, but not beyond.

In short, within the primates, many of the intellectual enhancements once considered uniquely hominin adaptations probably originated in the older and broader great ape lineage. Paleological evidence is consistent with a great ape grade of intelligence evolving with mid-Miocene hominids, as part and parcel of a biological package that includes larger brains, larger bodies, longer lives, and the mix of socioecological pressures the hominids faced and created (Russon & Begun, 2004). If so, these intellectual enhancements evolved as hominid adaptations to increasingly difficult life in moist tropical forests – not hominin adaptations to savanna life.

## The Intelligence of Early Humans

This section examines the archaeological evidence for the earliest indications of human intelligence and anthropological evidence for concurrent changes in the size and shape of the cranial cavity. It discusses the implications for the evolution of human intelligence.

### Homo Habilis

Ancestral humans started diverging from ancestral great apes approximately six million years ago. The first Homo lineage, *Homo habilis*, appeared approximately 2.4 million years ago in the Lower Paleolithic and persisted until 1.5 mya. The earliest known human inventions, referred to as *Oldowan* artifacts (after Olduvai Gorge, Tanzania, where they were first found), are widely attributed to *Homo habilis* (Semaw *et al*., 1997), although it is possible that they were also used by late australopithecenes (de Baune, 2004). They were simple, mostly single faced stone tools, pointed at one end (Leakey, 1971). These tools were most likely used to split fruits and nuts (de Baune, 2004), although some of the more recently constructed ones have sharp edges, and are found with cut-marked bones, suggesting that they were used to sharpen wood implements and butcher small game (Leakey, 1971; Bunn & Kroll, 1986).

Although these carefully planed and deliberately fashioned early tools are seen as marking a momentous breakthrough for our lineage, they were nevertheless simple and unspecialized; by our standards they were not indicative of a very flexible or creative kind of intelligence. The same tools were put to many uses instead of adapting them to precisely meet the task at hand. Mithen (1996) refers to minds at this time as possessing *generalized intelligence*, reflecting his belief that associative-level domain-general learning mechanisms, such as operant and Pavlovian conditioning, predominated. The minds of these early hominins have been referred to as *pre-representational*, because available artifacts show no indication that the hominins were capable of forming representations that deviated from their concrete sensory perceptions; their experience is considered to have been *episodic,* or tied to the present moment (Donald, 1993). Donald characterized their intelligence as governed by procedural memory. They could store perceptions of events and recall them in the presence of a reminder or cue, but they had little voluntary access to episodic memories without environmental cues. They were therefore unable to voluntarily shape, modify, or practice skills and actions, and they were unable to invent or refine complex gestures or means of communicating.

## The Massive Modularity Hypothesis

Evolutionary psychologists claim that the intelligence of *Homo* arose due to *massive modularity* (Buss, 1999, 2004; Buss *et al*., 1994; Cosmides & Tooby, 2002; Dunbar et al., 1994; Rozin, 1976; for an extensive critique see Buller, 2005 and Byrne, 2000). Cosmides and Tooby (1992) proposed that human intelligence evolved in the form of hundreds or thousands of functionally encapsulated (that is, not accessible to each other) cognitive modules. Each module was specialized to accomplish a specific task or solve a specific problem encountered by ancestral humans in their *environment of evolutionary adaptedness*, taken to be hunter-gatherer life in the Pleistocene. Modules for language, theory of mind, spatial relations, and tool use are among the modules proposed. These modules are supposedly content rich, pre-fitted with knowledge relevant to hunter-gatherer problems. It is also claimed that these modules exist today in more or less the same form as they existed in the Pleistocene, because too little time has passed for them to have undergone significant modification.

What is the current status of these ideas? Although the mind exhibits an intermediate degree of functional and anatomical modularity, neuroscience has not revealed vast numbers of hardwired, encapsulated, task-specific modules; indeed, the brain has been shown to be more highly subject to environmental influence than we thought (Wexler, 2006). Nevertheless, evolutionary psychology has made a valuable contribution by heightening awareness that the human mind is not an optimally designed machine; its structure and function reflect the pressures it was subjected to in over its long evolutionary history.

## Homo erectus

Approximately 1.9 million years ago, *Homo ergaster* and *Homo erectus* appeared, followed by archaic *Homo sapiens* and *Homo neanderthalensis*. The size of the *Homo erectus* brain was approximately 1,000 cc, about 25% larger than that of *Homo habilis*, at least twice as large as those of living great apes, and 75% the cranial capacity of modern humans (Aiello, 1996; Ruff *et al*., 1997; Lewin, 1999). *Homo erectus* exhibited many indications of enhanced ability to adapt to the environment to meet the demands of survival, including sophisticated, task-specific stone hand axes, complex stable seasonal home bases, and long-distance hunting strategies involving large game. By 1.6 mya, *Homo erectus* had dispersed as far as Southeast Asia, indicating the ability to adjust its lifestyle to different climates and habitats (Anton & Swisher, 2004; Cachel & Harris, 1995; Swisher, Curtis, Jacob, Getty, & Widiasmoro, 1994; Walker & Leakey, 1993). By 1.4 mya in Africa, West Asia, and Europe, *Homo erectus* had produced the Aschulean handaxe (Asfaw et al., 1992), a do-it-all tool that may have functioned as a social status symbol (Kohn & Mithen, 1999). The most notable characteristic of these tools is their biface (two-sided) symmetry. They probably required several stages of production, bifacial knapping, and considerable skill and spatial ability to achieve their final form.

Though anatomical evidence indicates the presence of Broca's area in the brain, suggesting that the capacity for language was present by this time (Wynn, 1998), verbal communication is thought to have been limited to (at best) pre-syntactical proto-language involving primarily short, nongrammatical utterances of one or two words (Dunbar, 1996). Mental processes during this time period probably strayed little from concrete sensory experience. The capacity for abstract thought, and for thinking about what one is thinking about (that is, metacognition), had not yet appeared.

## Social Explanations

There are multiple versions of the hypothesis that the origins of human intellect and onset of the archaeological record reflect a transition in cognitive or social abilities. *Homo erectus* were indeed probably the earliest humans to live in hunter-gatherer societies. One suggestion has been that they owe their achievements to onset of theory of mind (Mithen, 1998). However, as we have seen, there is evidence that other species possess theory of mind (Heyes, 1998), yet do not compare to modern humans in intelligence.

## Self-triggered Recall and Rehearsal Loop

Donald (1991) proposed that with the enlarged cranial capacity of *Homo erectus*, the human mind underwent the first of three transitions by which it – and the cultural matrix in which it is profoundly embedded – evolved from the ancestral, pre-hominin condition. Each transition entailed a new way of encoding representations in memory and storing them in collective memory so that they can later be drawn upon and shared with others.

This first transition is characterized by a shift from an *episodic* to a *mimetic mode* of cognitive functioning, made possible by onset of the capacity for voluntary retrieval of stored memories, independent of environmental cues. Donald refers to this as a "self-triggered recall and rehearsal loop." Self-triggered recall enabled hominins to access memories voluntarily and thereby act out[1] events that occurred in the past or that might occur in the future. Thus not only could the mimetic mind temporarily escape the here and now, but by miming or gesture, it could communicate similar escapes in other minds. The capacity to mime thus ushered forth what is referred to as a *mimetic* form of cognition and brought about a transition to the mimetic stage of human culture. The self-triggered recall and rehearsal loop also enabled hominins to engage in a stream of thought. One thought or idea evokes another, revised version of it, which evokes yet another, and so forth recursively. In this way, attention is directed away from the external world toward one's internal model of it. Finally, self-triggered recall allowed actors to take control over their own output, including voluntary rehearsal and refinement, and mimetic skills such as pantomime, reenactive play, self-reminding, imitative learning, and proto-teaching. In effect, it allows systematic evaluation and improvement of motor acts and adapting them to new situations, resulting in more refined skills and artifacts, and the capacity to use one's body as a communication device to act out events.

Donald's scenario becomes even more plausible in light of the structure and dynamics of associative memory (Gabora, 1998, 2003, 2007; Gabora & Aerts, 2009). Neurons are sensitive to *microfeatures* – primitive stimulus attributes such as a sound of a particular pitch, or a line of a particular orientation. Episodes etched in memory are *distributed* across a bundle or cell assembly of these neurons, and likewise, each neuron participates in the encoding of many episodes. Finally, memory is *content-addressable*, such that similar stimuli activate and get encoded in overlapping distributions of neurons. With larger brains, episodes are encoded in more detail, allowing for a transition from more coarse-grained to more fine-grained memory. Fine-grained memory means more microfeatures of episodes tend to be encoded, so there are more ways for distributions to overlap. Greater overlap meant more routes by which one memory can evoke another, making possible the onset of self-triggered recall and rehearsal, and paving the way for a more integrated internal model of the world, or worldview.

## Over a Million Years of Stasis

The handaxe persisted as the almost exclusive tool preserved in the archaeological record for over a million years, spreading by 500,000 years ago into Europe, where was it used until about 200,000 years ago. During this period, there was almost no change in tool design and little other evidence of new forms of intelligent behavior, with the exception of the first solid evidence for controlled use of fire, approximately 800,000 years ago (Goren-Inbar et al., 2004). There is, however, some evidence (such as charred animal bones at *Homo ergaster* sites) that fire may have been used substantially earlier.

### A **Second Increase in Brain Size**

Between 600,000 and 150,000 years ago there was a second spurt in brain enlargement (Aiello, 1996; Ruff *et al*., 1997), which marks the appearance of anatomically modern humans. It would make our story simple if the increase in brain size coincided with the burst of creativity in the Middle/Upper Paleolithic (Bickerton, 1990; Mithen, 1998), to be discussed shortly. But although *anatomically* modern humans had arrived, *behavioral* modernity had not. Leakey (1984) writes of anatomically modern human populations in the Middle East with little in the way of evidence for the kind of intelligence of modern humans and concludes, “The link between anatomy and behavior therefore seems to break” (p. 95). An exception to the overall lack of evidence for intellectual progress at this time is the advancement of the Levallois flake, which came into prominence approximately 250,000 years ago in the Neanderthal line. This suggests that cognitive processes were primarily first-order—tied to concrete sensory experience—rather than second-order—derivative, or abstract.

Perhaps this second spurt in encephalization exerted an impact on expressions of intelligence that left little trace in the archaeological record, such as ways of coping with increasing social complexity, or manipulating competitors (Baron-Cohen, 1995; Byrne & Whiten, 1988; Dunbar, 1996; Humphrey, 1976; Whiten, 1991; Whiten & Byrne, 1997; Wilson et al., 1996). Another possible reason for the apparent rift between anatomical and behavioral modernity is that while genetic changes necessary for cognitive modernity arose at this time, the fine-tuning of the nervous system to fully capitalize on these genetic changes took longer, or the necessary environmental conditions were not yet in place (Gabora, 2003). It is worth noting that other periods of revolutionary innovation, such as the Holocene transition to agriculture and the modern Industrial Revolution, occurred long after the biological changes that made them cognitively possible.

## The Spectacular Intelligence of Modern Humans

The European archaeological record indicates that an unparalleled transition occurred between 60,000 and 30,000 years ago at the onset of the Upper Paleolithic (Bar-Yosef, 1994; Klein, 1989a; Mellars, 1973, 1989a, 1989b; Soffer, 1994; Stringer & Gamble, 1993). Considering it “evidence of the modern human mind at work,” Richard Leakey (1984:93-94) writes: “unlike previous eras, when stasis dominated, ... [with] change being measured in millennia rather than hundreds of millennia.” Similarly, Mithen (1996) refers to the Upper Paleaolithic as the ‘big bang’ of human culture, exhibiting more innovation than in the previous six million years of human evolution.

At this time we see the more or less simultaneous appearance of traits considered diagnostic of behavioral modernity. They include the beginning of a more organized, strategic, season-specific style of hunting involving specific animals at specific sites, elaborate burial sites indicative of ritual and religion, evidence of dance, magic, and totemism, the colonization of Australia, and replacement of Levallois tool technology by blade cores in the Near East. In Europe, complex hearths and many forms of art appeared, including naturalistic cave paintings of animals, decorated tools and pottery, bone and antler tools with engraved designs, ivory statues of animals and sea shells, and personal decoration such as beads, pendants, and perforated animal teeth, many of which may have indicated social status (White 1989a,b). White (1982:176) also wrote of a "total restructuring of social relations". What is perhaps most impressive about this period is not the novelty of any particular artifact but that the overall pattern of change is cumulative; more recent artifacts resemble older ones but have modifications that enhance their appearance or functionality. This cumulative change is referred to as the *ratchet effect* (Tomasello, Kruger & Ratner, 1993), and some suggest it is uniquely human (Donald, 1998).

Whether this period was a genuine revolution culminating in behavioral modernity is hotly debated because claims to this effect are based on the European Palaeolithic record, and largely exclude the African record (McBrearty & Brooks, 2000); Henshilwood & Marean, 2003). Indeed, most of the artifacts associated with a rapid transition to behavioral modernity at 40,000–50,000 years ago in Europe are found in the African Middle Stone Age tens of thousands of years earlier. These artifacts include blades and microliths, bone tools, specialized hunting, long distance trade, art and decoration (McBrearty & Brooks, 2000), the Berekhat Ram figurine from Israel (d'Errico & Nowell, 2000), and an anthropomorphic figurine of quartzite from the Middle Ascheulian (ca. 400 ka) site of Tan-tan in Morocco (Bednark, 2003). Moreover, gradualist models of the evolution of cognitive modernity well before the Upper Palaeolithic find some support in archaeological data (Bahn, 1991; Harrold, 1992; Henshilwood & Marean, 2003; White, 1993; White *et al*., 2003). If modern human behaviors were indeed gradually assembled as early as 250,000–300,000 years ago, as McBrearty and Brooks (2000) argue, the transition falls more closely into alignment with the most recent spurt in human brain enlargement. However, the traditional and currently dominant view is that modern behavior appeared in anatomically modern humans in Africa between 50,000 and 40,000 years ago due to biologically evolved cognitive advantages, and that anatomically modern humans spread replacing existing species, including the Neanderthals in Europe e.g., Ambrose, 1998; Gamble, 1994; Klein, 2003; Stringer & Gamble, 1993). Thus, from this point onward, there was only one hominin species: the modern *Homo sapiens*.

Despite lack of overall increase in cranial capacity, the prefrontal cortex, and particularly the orbitofrontal region, increased disproportionately in size (Deacon, 1997; Dunbar, 1993; Jerison, 1973; Krasnegor, Lyon, & Goldman-Rakic, 1997; Rumbaugh, 1997) and it was likely a time of major neural reorganization (Henshilwood, d'Errico, Vanhaeren, van Niekerk, & Jacobs, 2000; Klein, 1999). These brain changes may have given rise to metacognition, or what Feist (2006) refers to as "meta-representational thought," that is, the ability to reflect on representations and think about thinking.

Whether or not it is considered a "revolution," it is accepted that the Middle/Upper Paleolithic was a period of unprecedented intellectual activity. How and why did it occur? Let us now review the most popular hypotheses for how and why behavioral modernity and its underlying intellectual capacities arose.

## Syntactic Language and Symbolic Reasoning

It has been suggested that at this time humans underwent a transition from a predominantly gestural to a vocal form of communication (Corballis, 2002). Although the ambiguity of the archaeological evidence means we may never know exactly when language began (Bednarik, 1992:30; Davidson & Noble, 1989), most scholars agree that earlier Homo and even Neanderthals may have been capable of primitive proto-language and the grammatical and syntactic aspects emerged at the start of the Upper Palaeolithic (Aiello & Dunbar, 1993; Bickerton, 1990, 1996; Dunbar, 1993, 1996; Tomasello, 1999).

Carstairs-McCarthy (1999) presented a modified version of this proposal, suggesting that although some form of syntax was present in the earliest languages, most of the later elaboration, including recursive embedding of syntactic structure, emerged in the Upper Paleolithic. Syntax enabled the capacity to state more precisely how elements are related and to embed them in other elements. Thus it enabled language to become general-purpose and applied in a variety of situations.

Deacon (1997) stresses that the onset of complex language reflects onset of the capacity to internally representing complex, abstract, internally coherent systems of meaning using symbols—items, such as words, that arbitrarily stand for other items, such as things in the world. The advent of language made possible what Donald (1991) refers to as the *mythic* or story-telling stage of human culture. It enhanced not just the ability to communicate with others, spread ideas from one individual to the next, and collaborate (thereby speeding up cultural innovation), but also the ability to think things through for oneself and manipulate ideas in a controlled, deliberate fashion (Reboul, 2007).

## Cognitive Fluidity, Connected Modules, and Cross-Domain Thinking

Another proposal is that the exceptional abilities exhibited by *Homo* in the Middle/Upper Paleolithic were due to the onset of *cognitive fluidity* (Fauconnier and Turner, 2002). Cognitive fluidity involves the capacity to draw analogies, to combine concepts and adapt ideas to new contexts, and to map across different knowledge systems, potentially employing multiple 'intelligences' simultaneously (Gardner, 1983; Langer, 1996; Mithen, 1996). Cognitive fluidity would have facilitated the weaving of experiences into stories, parables, and broader conceptual frameworks, and thereby the integration of knowledge and experience (Gabora & Aerts, 2009).

A related proposal has been put forward by Mithen (1996). Drawing on the evolutionary psychologist's notion of massive modularity, he suggests that the abilities of the modern human mind arose through the interconnecting of preexisting intellectual modules (that is, encapsulated or functionally isolated *specialized intelligences,* or cognitive domains) devoted to natural history, technology, social processes, and language. This interconnecting, he claims, is what enabled the onset of cognitive fluidity and allowed humans to map, explore, and transform conceptual spaces. Sperber (1994) proposed that the connecting of modules involved a special module, the "module of meta-representation," which contains "concepts of concepts" and enabled cross-domain thinking, and particularly analogies and metaphors.

## Contextual Focus: Shifting Between Explicit and Implicit Modes of Thought

These proposals for what kinds of cognitive change could have led to the Upper Paleolithic transition stress different aspects of cognitive modernity. Acknowledging a possible seed of truth in each, we begin to converge toward a common (if more complex) view. Concept combination is characteristic of *divergent thought*, which tends to be intuitive, diffuse, and associative.

Divergent thought is on the opposite end of the spectrum from the *convergent thought* stressed by Deacon, which tends to be logical, controlled, effortful, and reflective and symbolic. Converging evidence suggests that the modern mind engages in both (Arieti, 1976; Ashby & Ell, 2002; Freud, 1949; Guilford, 1950; James, 1890/1950; Johnson-Laird, 1983; Kris, 1952; Neisser, 1963; Piaget, 1926; Rips, 2001; Sloman, 1996; Stanovich & West, 2000; Werner, 1948; Wundt, 1896). This is sometimes referred to as the dual-process theory of human cognition (Chaiken & Trope, 1999; Evans & Frankish, 2009) and it is consistent with some current theories of cognition (Finke, Ward, & Smith, 1992; Gabora, 2000, 2002, 2003, under revision; S.B. Kaufman, this volume. Divergent processes are hypothesized to facilitate insight and idea generation, while convergent processes predominate during the refinement, implementation, and testing of an idea.

It has been proposed that the Paleolithic transition reflects genetic changes involved in the fine-tuning of the biochemical mechanisms underlying the capacity to shift between these modes of thought, depending on the situation, by varying the specificity of the activated cognitive receptive field (Gabora, 2003, 2007; for similar ideas see Howard-Jones & Murray, 2003; Martindale, 1995). This capacity is referred to as *contextual focus*[2] because it requires the ability to focus or defocus attention in response to the context or situation one is in. Defocused attention, by diffusely activating a broad region of memory, is conducive to divergent thought; it enables obscure (but potentially relevant) aspects of the situation to come into play. Focused attention is conducive to convergent thought; memory activation is constrained enough to hone in and perform logical mental operations on the most clearly relevant aspects. Note that contextual focus enables dynamic "resizing" of the activated brain region in response to the situation (as opposed to rigid compartmentalization).

Once the capacity to shrink or expand the field of attention came about, thereby improving the capacity to tailor one's mode of thought to the demands of the current situation, tasks requiring convergent thought (e.g., mathematical derivation), divergent thought (e.g., poetry), or both (e.g., technological invention) could be carried out more effectively. When the individual is fixated or stuck, and progress is not forthcoming, defocusing attention enables the individual to enter a more divergent mode of thought, and peripherally related elements of the situation begin to enter working memory until a potential solution is glimpsed. At this point attention becomes more focused, and thought becomes more convergent, as befits the fine-tuning of the idea and manifestation of it in the world.

Thus, the onset of contextual focus would have enabled hominins to adapt ideas to new contexts or combine them in new ways through divergent thought and fine-tune these unusual new combinations through convergent thought. In this way, the fruits of one mode of thought provide the ingredients for the other, culminating in a more fine-grained internal model of the world.

A related proposal is that this period marks the onset of the capacity to move between explicit and implicit modes of thought (Feist, 2007). Explicit thought involves the executive functions concerned with control of cognitive processes such as planning and decision making, while implicit thought encompasses the ability to automatically and nonconsciously detect complex regularities, contingencies, and covariances in our environment (Kaufman, DeYoung, Gray, Jiménez, Brown, & Mackintosh, N., under revision). A contributing factor to the emergence of the ability to shift between them may have been the expansion of the prefrontal cortex. This expansion probably enhanced the executive functions as well as the capacity to maintain and manipulate information in an active state in working memory. Indeed, individual differences in working memory capacity are strongly related to fluid intelligence in modern humans (Conway, Jarrold, Kane, & Miyake, 2007; Engle, Tuholski, Laughlin, & Conway, 1999; Kane, Hambrick, & Conway, 2005; Kaufman, DeYoung, Gray, Brown, & Mackintosh, 2009).

## Synthesizing the Various Accounts

The notion of mental modules amounts to an explicit compartmentalization of the brain for different tasks. However, this kind of division of labor – and the ensuing intelligence – would emerge unavoidably as the brain got larger *without* explicit high-level compartmentalization, due to the sparse, distributed, content-addressable manner in which neurons encode information (Gabora, 2003). Because neurons are tuned to respond to different microfeatures and a systematic relationship exists between the content of a stimulus and the distributed set of neurons that respond to it, neurons that respond to similar microfeatures are near one another (Churchland & Sejnowski, 1992; Smolensky, 1988). Thus, as the brain got larger and the number of neurons increased, and the brain accordingly responded to a greater variety of microfeatures, neighboring neurons tended to respond to microfeatures that were more similar, and distant neurons tended to respond to microfeatures that were more different. There were more ways in which distributed representations could overlap and new associations be made. Thus a weak modularity of sorts can emerge at the neuron level without any explicit compartmentalization going on, and it need not necessarily correspond to how humans carve up the world, that is, to categories such as natural history, technology, and so forth. Moreover, explicit connecting of modules is not necessary for new associations to be made; all that is necessary is that the relevant domains or modules be simultaneously accessible.

Let us return briefly to the question of why the burst of innovation in the Upper Paleolithic became apparent well after the second rapid increase in brain size approximately 500,000 years ago. A larger brain provided more room for episodes to be encoded, and particularly more association cortex for connections between episodes to be made, but it doesn't follow that this increased brain mass could straightaway be optimally navigated. It is reasonable that it took time for the anatomically modern brain to fine-tune how its components "talk" to each other such that different items could be merged or recursively revised and recoded in a coordinated manner (Gabora, 2003). Only then could the full potential of the large brain be realized. Thus the bottleneck may not have been sufficient brain size but sufficient sophistication in the *use of* the capacities that became available – for example, by way of contextual focus, or shifting between implicit and explicit thought.

# "Recent" Breakthroughs in the Evolution of Intelligence

Of course, the story of how human intelligence evolved does not end with the arrival of anatomical and behavioral modernity. The end of the ice age around 10,000–12,000 years ago witnessed the beginnings of agriculture and the invention of the wheel. Written languages developed around 5,000–6,000 years ago, and approximately 4,000 years ago astronomy and mathematics appear on the scene. We see the expression of philosophical ideas around 2,500 years ago, invention of the printing press 1,000 years ago, and the modern scientific method about 500 years ago. The past 100 years have yielded a technological explosion that has completely altered the daily routines of humans (as well as other species), the consequences of which remain to be seen. Donald (1991) claims that in recent time the abundance of new means of altering our environment and thereby creating an external, communally accessible form of memory brought about what he refers to as the *theoretic* stage of human cognitive.

# Why Did Intelligence Evolve?

We have examined how the *capacity* for human intelligence evolved over millions of years. We now address a fundamental question: *Why* did human intelligence evolve?

## Biological Explanations

We begin with biological explanations for the evolution of human intelligence. Biological explanations generally invoke natural selection as underlying the mechanism; that is, those who displayed a certain characteristic or behavior leave behind more offspring, or are "selected for." Thus, biological explanations have to do with competitive exclusion or "survival of the fittest." Because modifications that are acquired over the course of a lifetime – for example, through learning – do not get incorporated into the organism's genome or DNA, they are not inherited. Because they are not passed on to the next generation, they are not selected for. However, in some cases they may play an indirect role. We now look at a few of the factors that can influence what gets selected for, and thereby influence the evolution of intelligence.

### Intelligence as Evolutionary Spandrel

Some products of intelligence enhance survival and thus reproductive fitness. For example, the invention of weapons most likely evolved as an intelligent response to a need for protection from enemies and predators. For other expressions of intelligence, however, such as art, music, humor, fiction, religion, and philosophy, the link to survival and reproduction is not clear-cut. Why do we bother? One possibility is that art and so forth are not real adaptations but evolutionary spandrels: side-effects of abilities that evolved for other purposes (Pinker, 1997). Dennett argued that even language originally arose as an evolutionary spandrel.

### Group Selection

Even if intelligence is at least in part driven by individual-level biological selection forces, other forces may also be at work. Natural selection is believed to operate at multiple levels, including gene-level selection, individual-level election, sexual selection, kin selection, and group selection. Although there is evidence from archaeology, anthropology, and ethnography that individual-level selection plays a key role in human intelligence, other levels may have an impact as well.

### Sexual Selection

Some (e.g., Miller, 2000a,b) argue for a possible role of sexual selection in shaping intelligent behavior. According to the sexual-selection account, there is competition to mate with individuals who exhibit intelligence because it is (in theory) a reliable indicator of fitness. Intelligence may be the result of complex psychological adaptations whose primary functions were to attract mates, yielding reproductive rather than survival benefits. According to the "sexy-handaxe hypothesis" sexual selection pressures may have caused men to produce symmetric handaxes as a reliable indicator of cognitive, behavioral, and physiological fitness (Kohn, 1999; Kohn & Mithen, 1999). As Mithen (1996) noted, the symmetry of handaxes is attractive to the eye, but these tools require a huge investment in time and energy to make – a burden that makes their evolution difficult to account for in terms of strictly practical, survival purposes.

## The Baldwin Effect

Not all believe that the spandrel idea can account for the evolution of language. Pinker (1997) invoked the *Baldwin effect*. To understand how this works, note first that genetic diversity within a population is costly because if a superior trait exists, ideally all members of the population should converge on it. However, the advantage of genetic diversity comes to light in uncertain or changing environments; if one variant does not excel under the new conditions, another variant may. Baldwin's insight was that learning may increase the likelihood of evolutionary change by increasing behavioral flexibility, thereby reducing the evolutionary cost of genetic diversity. The idea is that if environmental uncertainty is being effectively dealt with at the *behavioral level,* it need no longer be looked after at the *genetic level*. Thus, although selective pressures cannot preserve the *results* of learning, they can act on any possible genetic factors underlying the *propensity* to learn.

The greater the proportion of individuals in a population who express themselves with language or use other kinds of symbols, the greater the value of language or symbol use to *other* individuals in this population. Therefore, natural selection can start to act on the genetic variation underlying the ability to learn. Individuals whose genetic makeup does *not* predispose them to use language or symbols are not selected for. In this way, the Baldwin effect provides a theoretically justifiable Darwinian explanation for evolution of the propensity to acquire language, use symbols, or indeed any trait whose complexity makes it difficult to see how it can be accounted for by orthodox natural selection.

According to Pinker, this is how the ability to learn language evolved. The Baldwin effect led to the evolution of a set of innate brain functions that (following Chomsky) he refers to as the *Language Acquisition Device*, or LAD. It is because the LAD is innate that there are developmental windows for language learning. This, he claims, is also the reason humans tend to learn language-typical sounds, words, and grammatical rules according to a stereotyped series of steps. Deacon (1997) also saw the Baldwin effect as playing an essential role in the evolution of human language, but in his account, acquisition of symbol use is emphasized much more than grammar.

Empirical proof that any particular facet of human intelligence can be accounted for by the Baldwin effect is difficult to obtain, but it does have computational support. Hinton and Nowlan (1987) ran a computer simulation using a "sexually reproducing" population of neural networks, which showed over generations a progressive increase in genes that enabled learning, accompanied by reduced genetic diversity (increased fixation). In other words, they provided computational evidence for the feasibility of the Baldwin effect.

## Cultural Explanations of Intelligence

The Baldwin effect predisposes us to face challenges and uncertainties through behavioral flexibility and learning (rather than exhibit hardwired diversity in the hopes that at least one of us will possess the right genes to meet whatever challenge comes along). It thus sets the stage for brain tissue that is relatively undifferentiated and adaptable, and subject to substantial modification through *other, nonbiological* influences such as culture. The drive to create is often compared with the drive to procreate, and evolutionary forces may be at the genesis of both. In other words, we may be tinkered with by two evolutionary forces: one that prompts us to act in ways that foster the proliferation of our biological lineage, and one that prompts us to act in ways that foster the proliferation of our cultural lineage. For example, it has been suggested that we exhibit a cultural form of altruism, such that we are kinder to those with whom we share ideas and values than to those with whom we share genes for eye color or blood type (Gabora, 1997). By contributing to the well-being of those who share our cultural makeup, we aid the proliferation of our "cultural selves." Similarly, when we are on the verge of an intellectual breakthrough, it may be that forces originating as part of cultural evolution are compelling us to give all we have to our ideas and thereby impact our cultural lineage, much as biological forces compel us to provide for our children.

It has been proposed that the evolution of ideas through culture works in a manner akin to the evolution of the earliest life forms ((Gabora, 1998, 2000, 2004, 2008; Gabora & Aerts, 2009). Recent work indicates that early life emerged and replicated through a self-organized process referred to as *autocatalysis*, in which a set of molecules catalyzes (speeds up) the reactions that generate other molecules in the set, until as a whole they self-replicate (Kaufman, 1993). Such a structure is *self-regenerating* because the whole is reconstituted through the interactions of the parts (Maturana & Varela, 1980). These earliest precursors of life evolved not through natural selection and competitive exclusion or "survival of the fittest," like present-day life, but rather by transformation and communal exchange (Gabora, 2006; Vetsigian et al., 2006). Because replication of these pre-DNA life forms occurred through regeneration of catalytic molecules rather than (as with present-day life) by using a genetic self-assembly code, acquired traits were inherited. In other words, their evolution was, like that of culture, Lamarckian.

This suggests that it is worldviews that evolve through culture, through the same non-Darwinian process as the earliest forms of life evolved, and products of our intelligence such as tools and architectural plans are external manifestations of this process; they reflect the states of the particular worldviews that generate them. The idea is that like these early life forms, worldviews evolve not through natural selection but through self-organization and communal exchange of innovations. One does not accumulate elements of culture transmitted from others like items on a grocery list but hones them into a unique tapestry of understanding, a worldview, which like these early life forms is autopoietic in that the whole emerges through interactions among the parts. It is *self-mending* in the sense that, just as injury to the body spontaneously evokes physiological changes that bring about healing, events that are problematic or surprising or evoke cognitive dissonance spontaneously evokes streams of thought that attempt to generate an intelligent solution to the problem or reconcile the dissonance (Gabora, 1999). Thus it is proposed that what fuels intelligent thought is the self-organizing, self-mending nature of a worldview.

## Conclusions

This chapter began with an overview of the primate context out of which human intelligence emerged, concentrating on the modern great apes. Modern great apes offer the best and indeed the only living models of the cognitive platform from which human intelligence evolved. The cognitive abilities that great apes demonstrate suggest that a more sophisticated intelligence predated the human lineage than we have traditionally believed. Many of the intellectual qualities believed to have evolved in early *Homo* are now recognized in the great apes – including basic symbolic cognition, creativity, and cultural transmission – so they most likely evolved in ancestral great apes of the mid-Miocene era, well before the hominins diverged. The evolutionary changes proposed to have culminated in modern human intelligence may remain correct, but when and where they occurred and what the archaeological record implies about hominin intelligences may need to be reconsidered.

We continued to a brief tour of the history of *Homo sapiens*, starting six million years ago when we began diverging from ancestral large apes. The earliest signs of creativity in *Homo* are simple stone tools, thought to be made by *Homo habilis*, just over two million years ago. Though primitive, they marked a momentous breakthrough: the arrival of a species within our own lineage that would eventually refashion to its liking an entire planet. With the arrival of *Homo erectus* roughly 1.8 million years ago, there was a dramatic enlargement in cranial capacity coinciding with solid evidence of enhanced intelligence: task-specific stone handaxes, complex stable seasonal habitats, and signs of coordinated, long-distance hunting. The larger brain may have allowed items encoded in memory to be more fine-grained, which facilitated the forging of richer associations between them, and paved the way for self-triggered thought and rehearsal and refinement of skills, and thus the ability mentally go beyond "what is" to "what could be."

Another rapid increase in cranial capacity occurred between 600,000 and 150,000 years ago. It preceded by some hundreds of thousands of years the sudden flourishing of human-made artifacts between 60,000 and 30,000 years ago in the Middle/Upper Paleolithic, which is associated with the beginnings of art, science, politics, religion, and probably syntactical language. The time lag suggests that behavioral modernity arose due not to new brain parts or increased memory but to a more sophisticated way of *using* memory, which may have involved the enhancement of symbolic thinking, cognitive fluidity, and the capacity to shift between convergent and divergent or explicit and implicit modes of thought. Also, the emergence of meta-cognition enabled our ancestors to reflect on and even override their own nature.

The breadth of material that must be weighed to reconstruct models of how and why human intelligence evolved is vast, ranging from characterizations of modern human intelligence and brains to inferring ancestral intelligences from the fragmentary evidence available, identifying and weighing how ecological and social pressures may have guided evolutionary change, and reconstructing when and where these changes occurred. As we continue to study, our understanding of these factors continues to change. An important task facing us now is adjusting views that were built on evidence from within the *Homo* lineage in light of evidence on the hominid lineage from which *Homo* evolved – especially, evidence of greater similarities between humans and great apes in intelligence than traditionally believed.

The striking pattern that emerges from juxtaposing these two perspectives is a disjunction: Based on comparing great apes' tool use with *Homo* tool artifacts, for instance, living great apes show evidence of intellectual capabilities that resemble those inferred in early *Homo* (Byrne, 2004). Great apes' ancestors from the mid-late Miocene had brains of comparable size, so these intellectual capabilities may have been potentiated as early as 12–14 mya (Begun & Kordos, 2004). One implication is that a grade of intelligence that generates basic symbolism and creativity evolved as an adaptation to forested environments of Eurasia during the Miocene, not much more recent savanna habitats in East Africa. If hominids could evolve larger brains and enhanced intelligence, why did they stop at moderate enhancements? A good guess is that they never really got away from fruit diets and this may have limited their capacity to take in enough energy to enlarge their brains more. If so, what ancestral hominins' mix of social and ecological pressures (e.g., savanna life, eating more meat) enabled was evolutionary enlargement of hominid brains, which enabled elaborations to hominid intelligence. The intellectual advances that evolved with *Homo* may have been higher level, not basic, symbolism – possibly, symbol systems. These hominin elaborations beyond great ape intelligence are what need evolutionary explanation, and they make better sense in light of great apes' grade of intelligence and its evolutionary history.

This chapter also addressed the question, at some level, of *why* human intelligence evolved, and whether it is still evolving. Several biological explanations for the evolution of intelligence have been proposed. One is that certain of its expressions emerged as evolutionary spandrels. Sexual selection, group selection, and the Baldwin effect have also been implicated as playing a role in shaping intelligence. Another possibility derives from the theory that culture constitutes a second form of evolution, and that our thought and behavior are shaped by *two* distinct evolutionary forces. Just as the drive to procreate ensures that at least some of us make a dent in our biological lineage, the drive to create may enable us to make a dent in our cultural lineage. It was noted that the self-organized, self-regenerating autocatalytic structures widely believed to be the earliest forms of life did not evolve through natural selection either, but through a Lamarckian process involving communal exchange of innovations. It has been proposed that what evolves through culture is individuals' internal models of the world, or worldviews, and that like early life they are self-organized and self-regenerating. They evolve not through survival of the fittest but through transformation. By understanding the evolutionary origins of human intelligence, we gain perspective on pressing issues of today and are in a better position to use our intelligence to direct the future course of our species and our planet.

## Acknowledgments

This work was funded in part by grants to the first author from the Social Sciences and Humanities Research Council of Canada (SSHRC) and the GOA Project of the Free University of Brussels, and grants to the second author from the Natural Sciences and Engineering Research Council of Canada, the LSB Leakey Foundation, and York University.

## References

Aiello, L. C. (1996). Hominine preadaptations for language and cognition. In P. Mellars & K. Gibson (Eds.), *Modeling the early human mind* (pp. 89–99). Cambridge, UK: McDonald Institute Monographs.

Aiello, L. C., & Dunbar, R. (1993). Neocortex size, group size, and the evolution of language. *Current Anthropology, 34,* 184–193.

Ambrose, S. H. (1998). Chronology of the later stone age and food production in East Africa. *Journal of Archaeological Science, 25,* 377–392.

Antón, S. C., & Swisher, C. C. (2004). Early dispersals of homo from Africa. *Annual Review of Anthropology, 33,* 271–296.

Arieti, S. (1976). *Creativity: The magic synthesis*. New York, NY: Basic Books.

Asfaw, B., Yonas, B., Gen, S., Walterm R. C., White, T. D., et al. (1992). The earliest acheulean from konso-gardula. *Nature, 360*, 732–735.

Ashby, F. G., & Ell, S. W. (2002). Single versus multiple systems of learning and memory. In J. Wixted & H. Pashler (Eds.), *Stevens' handbook of experimental psychology: Vol. 4. Methodology in experimental psychology*. New York, NY: Wiley.

Aunger, R. (2000). *Darwinizing culture*. Oxford, UK: Oxford University Press.

Bahn, P. G. (1991). Pleistocene images outside Europe. *Proceedings of the Prehistoric Society, 57,* 99–102.

J. H. Barkow, L. Cosmides, & J. Tooby, Eds. (1992). *The adapted mind: Evolutionary psychology and the generation of culture*. New York: Oxford University Press.

Bar-Yosef, O. 1994. The contribution of southwest Asia to the study of the origin of modern humans. In M. Nitecki & D. Nitecki (Eds.) Origins of anatomically modern humans. Plenum Press.

Bar-Yosef, O., Vandermeersch, B., Arensburg, B., Goldberg, P., & Laville, H. (1986). New data on the origin of modern an in the Levant. *Current Anthropology, 27,* 63–64.

Beck, B. B. (1980). *Animal tool behavior: The use and manufacture of tools by animals*. New York, NY: Garland STPM Press.

Bednarik, R. G. (1992). Paleoart and archaeological myths. *Cambridge Archaeological Journal, 2*, 27–57.

Bednarik, R. G. (2003). A figurine from the African Acheulian. *Current Anthropology, 44*, 405–413.

Begun, D. R., & Kordos, L. (2004). Cranial evidence and the evolution of intelligence in fossil apes. In A. E. Russon & D. R. Begun (Eds.), *The evolution of thought: Evolutionary origins of great ape intelligence* (pp. 260–279). Cambridge, UK: Cambridge University Press.

Bentley, R. A., Hahn, M. W., & Shennan, S. J. (2004). Random drift and culture change. *Proceedings of the Royal Society: Biology, 271*, 1443–1450.

Bickerton, D. (1990). *Language and species.* Chicago: Chicago University Press.

Bickerton, D. (1996). *Language and human behavior*. London: UCL Press.

Blackmore, S. J. (1999). *The meme machine.* Oxford: Oxford University Press.

Blake, J. (2004). Gestural communication in the great apes. In A. E. Russon & D. R. Begun (Eds.), *The evolution of thought: Evolutionary origins of great ape intelligence* (pp. 61–75). Cambridge, UK: Cambridge University Press.

Boden, M. (1990). *The creative mind: Myths and mechanisms*. Grand Bay, NB: Cardinal.

Boesch, C. (1991). Teaching in wild chimpanzees. *Animal Behaviour*, *41*, 530–532.

Boesch, C. (1996). Three approaches for assessing chimpanzee culture. In A. E. Russon, K. A. Bard, & S. T. Parker (Eds.), *Reaching into thought: The minds of the great apes* (pp. 404–429). Cambridge, UK: Cambridge University Press.

Boesch, C., & Boesch-Achermann, H. (2000). *The chimpanzees of the Taï Forest: Behavioural ecology and evolution.* Oxford, UK: Oxford University Press.

Boyd, R., & Richerson, P. (1985). *Culture and the evolutionary process*. Chicago, IL: University of Chicago Press.

Boysen, S. T., Berntson, G. G., Hannan, M. B., & Cacioppo, J. T. (1996). Quantity-based inference and symbolic representations in chimpanzees (*Pan troglodytes*). *Journal of Experimental Psychology: Animal Behavior Processes*, *22*, 76–86.

Buller, D. J. (2005). *Adapting minds*. Cambridge, MA: MIT Press.

Bunn, H. T., & Kroll, E. M. (1986). Systematic butchery by plio/pleistocene hominids at Olduvai Gorge, Tanzania. *Current Anthropology, 27,* 431–452.

Buss, D. M. (1994). *The evolution of desire: Strategies of human mating*. New York, NY: Basic Books.

Buss, D. M. (1999/2004). *Evolutionary psychology: The new science of the mind*. Boston, MA: Pearson.

Byrne, R. W. (1995). *The thinking ape*. Oxford, UK: Oxford University Press.

Byrne, R. W. (2004). The manual skills and cognition that lie behind hominid tool use. In A. E. Russon & D. R. Begun (Eds.), *The evolution of thought: Evolutionary origins of great ape intelligence* (pp. 31–44). Cambridge, UK: Cambridge University Press.

Byrne, R. W., & Russon, A. E. (1998). Learning by imitation: A hierarchical approach. *Behavioural and Brain Sciences*, *21*, 667–721.
Byrne, R. W., & Whiten, A. (1990). Tactical deception in primates: The 1990 database. *Primate Report, 27*, 1–101.
Byrne, R. W., & Whiten, A. (Eds.). (1988). *Machiavellian intelligence: Social expertise and the evolution of intellect in monkeys, apes, and humans*. Oxford, UK: Clarendon Press.
Cachel, S., & Harris, J. W. K. (1995). Ranging patterns, land-use and subsistence in homo erectus from the perspective of evolutionary ecology. In J. R. F. Bower & S. Sartono (Eds.), *Evolution and ecology of homo erectus* (pp. 51–66). Leiden, the Netherlands: Pithecanthropus Centennial Foundation.
Carlson, S. M., Davis, A. C., & Leach, J. G. (2005). Less is more: Executive function and symbolic representation in preschool children. *Psychological Science*, 16, 609–616.
Carstairs-McCarthy, A. (1999). *The origins of complex language*. Oxford, UK: Oxford University Press.
Cavalli-Sforza, L. L., & Feldman, M. W. (1981). *Cultural transmission and evolution: A quantitative approach*. Princeton, NJ: Princeton University Press.
Chaiken, S., & Trope, Y. (1999). *Dual-process theories in social psychology*. New York, NY: Guilford Press.
Churchland, P. S. & T. Sejnowski (1992). *The computational brain*. Cambridge, MA: MIT Press.
Conway, A. R. A., Jarrold, C., Kane, M. J., Miyake, A., & Towse, J. N. (2007). *Variation in working memory*. New York, NY: Oxford University Press.
Corballis, M. (2002). *From hand to mouth: The origins of language.* Princeton, NJ: Princeton University Press.
Cosmides, L., & Tooby, J. (1992). Cognitive adaptations for social exchange. In J. Barkow, L. Cosmides, & J. Tooby (Eds.), *The adapted mind* (pp. 163–228). New York, NY: Oxford University Press.
Darwin, C. (1871). *The descent of man, and selection in relation to sex* (2 vols.). London, UK: John Murray.
Davidson, I., & Noble, W. (1989) The archaeology of perception: Traces of depiction and language. *Current Anthropology, 30*(2), 125–155.
Dawkins, R. (1975). *The selfish gene*. Oxford, UK: Oxford University Press.
Deacon, T. W. (1997). *The symbolic species: The coevolution of language and the brain.* New York, NY: W.W. Norton.
de Beaune, S. A. (2004). The invention of technology: Prehistory and cognition. *Current Anthropology, 45,* 139–162.
Dennett, D. (1976). Conditions of personhood. In A. Rorty (Ed.), *The identities of persons* (pp. 175–197). Berkeley: University of California Press.
Dennett, D. (1995). *Darwin's dangerous idea: Evolution and the meaning of life*. New York, NY: Simon & Schuster.
D'Errico, F., & Nowell, A. (2000). A new look at the berekhat ram figurine: Implications for the origins of symbolism. *Cambridge Archaeological Journal, 10,* 123–167.
Donald, M. (1991). *Origins of the modern mind: Three stages in the evolution of culture and cognition*. Cambridge, MA: Harvard University Press.
Donald, M. (1998). Hominid enculturation and cognitive evolution. In Colin Renfrew & C. Scarre (Eds.), *Cognition and material culture: The archaeology of symbolic storage* (pp. 7–17). McDonald Institute Monographs.

Dugatkin, L. A. (2001). *Imitation factor: Imitation in animals and the origin of human culture.* New York, NY: Free Press.
Dunbar, R. (1993). Coevolution of neocortical size, group size, and language in humans. *Behavioral and Brain Sciences, 16*(4), 681–735.
Dunbar, R. (1996). *Grooming, gossip, and the evolution of language*. London, UK: Faber & Faber.
Durham, W. (1991). *Coevolution: Genes, culture, and human diversity*. Stanford, CA: Stanford University Press.
Emery, N. J., & Clayton, N. S. (2004). The mentality of crows: Convergent evolution of intelligence in corvids and apes. *Science*, *306*, 1903–1907.
Engle, R. W., Tuholski, S. W., Laughlin, J. E., & Conway, A. R. A. (1999). Working memory, short-term memory and general fluid intelligence: A latent variable approach. *Journal of Experimental Psychology: General, 128*, 309–331.
Evans, J., & Frankish, K. (2009). *In two minds: Dual processes and beyond.* New York, NY: Oxford University Press.
Fauconnier, G., & Turner, M. (2002). *The way we think: Conceptual blending and the mind's hidden complexities.* New York, NY: Basic Books.
Feist, G. (2006). *The psychology of science and the origins of the scientific mind*. New Haven, CT: Yale University Press.
Finke, R. A., Ward, T. B., & Smith, S. M. (1992). *Creative cognition: Theory, research, and applications*. Cambridge, MA: MIT Press.
French, R. (1994). *Ancient natural history*. London, UK: Routledge.
Freud, S. (1949). *An outline of psychoanalysis*. New York, NY: W. W. Norton.
Gabora, L. (1995). Meme and variations: A computer model of cultural evolution. In L. Nadel & D. Stein (Eds.), *Lectures in complex systems* (pp. 471–486). Reading, MA: Addison-Wesley.
Gabora, L. (1997). The origin and evolution of culture and creativity. *Journal of Memetics: Evolutionary Models of Information Transmission, 1*(1).
Gabora, L. (1998). Autocatalytic closure in a cognitive system: A tentative scenario for the origin of culture. *Psycoloquy, 9*(67).
Gabora, L. (1999). Weaving, bending, patching, mending the fabric of reality: A cognitive science perspective on worldview inconsistency. *Foundations of Science, 3*(2), 395–428.
Gabora, L. (2000). Conceptual closure: Weaving memories into an interconnected worldview. In G. Van de Vijver & J. Chandler (Eds.), *Closure: Emergent organizations and their dynamics*. New York, NY: Annals of the New York Academy of Sciences.
Gabora, L. (2003). Contextual focus: A tentative cognitive explanation for the cultural transition of the middle/upper Paleolithic. In R. Alterman & D. Hirsch (Eds.), *Proceedings of the 25th annual meeting of the Cognitive Science Society.* Boston. MA: Erlbaum.
Gabora, L. (2004). Ideas are not replicators but minds are. *Biology & Philosophy, 19*(1), 127–143.
Gabora, L. (2006). Self-other organization: Why early life did not evolve through natural selection. *Journal of Theoretical Biology, 241*(3), 443–450.
Gabora, L. (2008). The cultural evolution of socially situated cognition. *Cognitive Systems Research*, *9*(1–2), 104–113.
Gabora L & Aerts D. (2009). A model of the emergence and evolution of integrated worldviews. *Journal of Mathematical Psychology, 53, 434*-451.

Gabora, L. (2010). Revenge of the 'neurds': Characterizing creative thought in terms of the structure and dynamics of human memory. *Creativity Research Journal, 22*(1), 1-13.

Gamble, C. (1994). *Timewalkers: The prehistory of global colonization.* Cambridge, MA: Harvard University Press.

Gardner, H. (1983). *Frames of mind: The theory of multiple intelligences*. New York, NY: Basic Books.

Gardner, H. (1993). *Creating minds: An anatomy of creativity seen through the lives of Freud, Einstein, Picasso, Stravinsky, Eliot, Graham and Gandhi*. New York, NY: Basic Books.

Gómez, J. C. (2004). *Apes, monkeys, children, and the growth of mind.* Cambridge, MA: Harvard University Press.

Goodall, J. (1963). My life among wild chimpanzees. *National Geographic*, *124*, 272–308.

Goodall, J. (1986). *The chimpanzees of Gombe: Patterns of behavior*. Cambridge, MA: Harvard University Press.

Goren-Inbar, N., Alperson, N., Kislev, M. E., Simchoni, O., & Melamed., Y. (2004). Evidence of Hominin control of fire at Gesher Benot Ya'aqov, Israel. *Science, 304,* 725–727.

Guilford, P. J. (1950). Creativity. *American Psychologist, 5,* 444–454.

Harrold, F. (1992.) Paleolithic archaeology, ancient behavior, and the transition to modern Homo. In G. Bräuer & F. Smith (Eds.), *Continuity or replacement: Controversies in Homo sapiens evolution* (pp. 219–30). Rotterdam: Balkema.

Henshilwood, C., d'Errico, F., Vanhaeren, M., van Niekerk, K., & Jacobs, Z. (2004). Middle stone age shell beads from South Africa, *Science, 304,* 404.

Henshilwood, C. S., & Marean, C. W. (2003). The origin of modern human behavior. *Current Anthropology, 44,* 627–651.

Heyes, C. M. (1998). Theory of mind in nonhuman primates. *Behavioral and Brain Sciences, 211*, 104–134.

Hinton, G. E. & Nowlan, S. J. (1987). How learning can guide evolution. *Complex Systems, 1,* 495-502.

Hirata, S., & Fuwa, K. (2007). Chimpanzees (*Pan troglodytes*) learn to act with other individuals in a cooperative task. *Primates*, *48*, 13–21.

Hof, P. R., Chanis, R., & Marino, L. (2005). Cortical complexity in cetacean brains. *Anatomical Record Part A*, *287a,* 1142–1152.

Howard-Jones, P.A., & Murray, S. (2003). Ideational productivity, focus of attention, and context. *Creativity Research Journal, 15*(2&3), 153–166.

Howes, J. M. A. (1999). Prodigies and creativity. In R. J. Sternberg (Ed.), *Handbook of creativity*. Cambridge, UK: Cambridge University Press.

Humphrey, N. (1976). The social function of intellect. In P. P. G. Bateson & R. A. Hinde (Eds.), *Growing points in ethology* (pp. 303–317). Cambridge, UK: Cambridge University Press.

Jablonka, E., & Lamb, M. (2005). *Evolution in four dimensions: Genetic, epigenetic, behavioural and symbolic variation in the history of life.* Cambridge MA: MIT Press.

James, W. (1890/1950). *The principles of psychology.* New York, NY: Dover.

Jerison, H. J. (1973). *Evolution of the brain and intelligence.* New York, NY: Academic Press.

Johnson-Laird, P. N. (1983). *Mental models*. Cambridge, MA: Harvard University Press.

Kane, M. J., Hambrick, D. Z., & Conway, A. R. A. (2005). Working memory capacity and fluid intelligence are strongly related constructs: Comment on Ackerman, Beier, and Boyle. *Psychological Bulletin, 131*, 66–71.

Kauffman, S. (1993). *Origins of order*. New York, NY: Oxford University Press.

Kaufman, S. B., DeYoung, C. G., Gray, J. R., Brown, J., & Mackintosh, N. (2009). Associative learning predicts intelligence above and beyond working memory and processing speed. *Intelligence*. <AU: Please add vol. and page numbers>

Kaufman, S. B., DeYoung, C. G., Gray, J. R., Jiménez, L., Brown, J., & Mackintosh, N. (under revision). *Implicit learning as an ability*.

Klein, R. G. (1989). Biological and behavioral perspectives on modern human origins in South Africa. In P. Mellars & C. Stringe (Eds.), *The human revolution.* Edinburgh, UK: Edinburgh University Press.

Klein, R. G. (1999). *The human career: Human biological and cultural origins.* Chicago, IL: University of Chicago Press.

Klein, R. G. (2003). Whither the Neanderthals? *Science, 299,* 1525–1527.

Kohn, M. (1999). A race apart. *Index on Censorship, 28*(3), 79.

Kohn, M., & Mithen, S. (1999). Handaxes: Products of sexual selection? *Antiquity, 73*, 281.

Krasnegor, N., Lyon, G. R., & Goldman-Rakic, P. S. (1997). *Prefrontal cortex: Evolution, development, and behavioral neuroscience.* Baltimore, MD: Brooke.

Kris, E. (1952). *Psychoanalytic explorations in art*. New York, NY: International Universities Press.

Kuhlmeier, V. A., Boysen, S. T., & Mukobi, K. L. (1999). Scale-model comprehension by chimpanzees (*Pan troglodytes*). *Journal of Comparative Psychology*, *113*, 396–402.

Langer, J. (1996). Heterochrony and the evolution of primate cognitive development. In A. E. Russon, K. A. Bard, & S. T. Parker (Eds.), *Reaching into thought: The minds of the great apes* (pp. 257–277). Cambridge, UK: Cambridge University Press.

Leakey, M. D. (1971). *Olduvai gorge: Excavations in beds I and II, 1960–1963.* Cambridge, UK: Cambridge University Press.

Leakey, R. (1984). *The origins of humankind*. New York, NY: Science Masters Basic Books.

Leijnen, Gabora, & von Ghyczy. (in press). <AU: Please give first initials for authors>Is it better to invent or imitate? A computer simulation. *International Journal of Software and Informatics*.

MacLeod, C. (2004). What's in a brain? The question of a distinct brain anatomy in great apes. In A. E. Russon & D. R. Begun (Eds.), *The evolution of thought: Evolutionary origins of great ape intelligence* (pp. 105–121). Cambridge, UK: Cambridge University Press.

Martindale, C. (1995). Creativity and connectionism. In S. M. Smith, T. B. Ward, & R. A. Finke (Eds.), *The creative cognition approach* (pp. 249–268). Cambridge MA: MIT Press.

Matsuzawa, T. (1991). Nesting cups and metatools in chimpanzees. *Behavioral and Brain Sciences,* 14(4), 570–571.

Matsuzawa, T. (2001). Primate foundations of human intelligence: A view of tool use in nonhuman primates and fossil hominids. In T. Matsuzawa (Ed.), *Primate origins of human cognition and behavior* (pp. 3–25). Tokyo: Springer-Verlag.

Matsuzawa, T., Tomonaga, M., & Tanaka, M. (Eds.). (2006). *Cognitive development in chimpanzees*. Tokyo: Springer.

Maturana, R. H., & Varela, F. J. (1980). *Autopoiesis and cognition: The realization of the living.* New York, NY: Springer.

McBrearty, S., & Brooks, A. S. (2000). The revolution that wasn't: A new interpretation of the origin of modern human behavior. *Journal of Human Evolution, 39*, 453–563.

Mellars, P. (1973). The character of the middle-upper transition in South-West France. In C. Renfrew (Eds.), *The explanation of culture change.* London, UK: Duckworth.

Mellars, P. (1989a). Technological changes in the middle-upper Paleolithic transition: Economic, social, and cognitive perspectives. In P. Mellars & C. Stringer (Eds.), *The human revolution.* Edinburgh, UK: Edinburgh University Press.

Mellars, P. (1989b). Major issues in the emergence of modern humans. *Current Anthropology, 30*, 349–385.

Miles, H. L., Mitchell, R. W., & Harper, S. (1996). Simon says: The development of imitation in an enculturated orangutan. In A. E. Russon, K. A. Bard, & S. T. Parker (Eds.), *Reaching into thought: The minds of the great apes* (pp. 278–299). Cambridge, UK: Cambridge University Press.

Miller, G. F. (2000a). *The mating mind: How sexual choice shaped the evolution of human nature.* London, UK: Vintage.

Miller, G. F. (2000b). Sexual selection for indicators of intelligence. *Novartis Foundation Symposium, 233*, 260–270; discussion 270–280.

Mithen, S. (1996). *The prehistory of the mind: The cognitive origins of art and science*. London, UK: Thames and Hudson.

Mithen, S. (Ed.). (1998). *Creativity in human evolution and prehistory*. London, UK: Routledge.

Neisser, U. (1963). The multiplicity of thought. *British Journal of Psychology*, *54,* 1–14.

Newman, S. A. & Müller, G. B. (1999). Morphological evolution: Epigenetic mechanisms. In *Embryonic encyclopedia of life sciences*. London, UK: Nature Publishing Group.

Parker, S. T. (1996). Apprenticeship in tool-mediated extractive foraging: The origins of imitation, teaching, and self-awareness in great apes. In A. E. Russon, K. A. Bard, & S. T. Parker (Eds.), *Reaching into thought: The minds of the great apes* (pp. 348–370). Cambridge, UK: Cambridge University Press.

Parker, S. T., & Gibson, K. R. (Eds.). (1990). *"Language" and intelligence in monkeys and apes: Comparative developmental perspectives*. Cambridge, UK: Cambridge University Press.

Parker, S. T., & McKinney, M. (1999). *Origins of intelligence: The evolution of cognitive development in monkeys, apes, and humans*. Baltimore, MD: Johns Hopkins University Press.

Piaget, J. (1926). *The language and thought of the child*. Kent, UK: Harcourt Brace.

Pinker, S. (1997). *How the mind works*. New York, NY: W. W. Norton.

Potts, R. (2004). Paleoenvironments and the evolution of adaptability in great apes. In A. E. Russon & D. R. Begun (Eds.), *The evolution of thought: Evolutionary origins of great ape intelligence* (pp. 237–259). Cambridge, UK: Cambridge University Press.

Povinelli, D. (2000). *Folk physics for apes: The chimpanzee's theory of how the world works.* New York, NY: Oxford University Press.

Premack, D. (1988). "Does the chimpanzee have a theory of mind?" revisited. In R. W. Byrne & A. Whiten (Eds.), *Machiavellian intelligence: Social expertise and the evolution of intellect in monkeys, apes and humans* (pp. 160–179). Oxford, UK: Oxford University Press.

Premack, D., & Woodruff, G. (1978). Does the chimpanzee have a theory of mind? *Behavioral and Brain Sciences, 1*, 515–526.

Reader, S. M., & Laland, K. N. (Eds.). (2003). *Animal innovation*. Oxford, UK: Oxford University Press.

Reboul, A. (2007). Does the Gricean distinction between natural and non-natural meaning exhaustively account for all instances of communication? *Pragmatics & Cognition, 15*(2), 253–276.

Rips, L. (2001). Necessity and natural categories. *Psychological Bulletin, 127*(6), 827–852.

Rosch, R. H. (1975). Cognitive reference points. *Cognitive Psychology, 7,* 532–47.

Rozin, P. (1976). The evolution of intelligence and access to the cognitive unconscious. In J. M. Sprague & A. N. Epstein (Eds.), *Progress in psychobiology and physiological psychology*. New York, NY: Academic Press.

Ruff, C., Trinkaus, E., & Holliday, T. (1997). Body mass and encephalization in Pleistocene Homo. *Nature, 387,* 173–176.

Rumbaugh, D. M. (1997). Competence, cortex, and primate models: A comparative primate perspective. In N. A. Krasnegor, G. R. Lyon, & P. S. Goldman-Rakic (Eds.), *Development of the prefrontal cortex: Evolution, neurobiology, and behavior* (pp. 117–139). Baltimore, MD: Paul.

Rumbaugh, D. M., & Washburn, D. A. (2003). *Intelligence of apes and other rational beings*. New Haven, CT: Yale University Press.

Russon A. E. (1998). The nature and evolution of intelligence in orangutans (Pongo pygmaeus). *Primates*, *39*, 485–503.

Russon, A. E. (1999). Naturalistic approaches to orangutan intelligence and the question of enculturation. *International Journal of Comparative Psychology*, *12*, 181–202.

Russon, A. E. (2002). Pretending in free-ranging rehabilitant orangutans. In R. W. Mitchell (Ed.), *Pretending and imagination in animals and children* (pp. 229–240). Cambridge, UK: Cambridge University Press.

Russon, A. E. (2003). Innovation and creativity in forest-living rehabilitant orangutans. In S. M. Reader & K. N. Laland (Eds.), *Animal innovation* (pp. 279–306). Oxford, UK: Oxford University Press.

Russon, A. E. (2004). Great ape cognitive systems. In A. E. Russon & D. R. Begun (Eds.), *The evolution of thought: Evolutionary origins of great ape intelligence* (pp. 76–100). Cambridge, UK: Cambridge University Press.

Russon, A. E., Bard, K. A., & Parker, S. T. (Eds.). (1996). *Reaching into thought: The minds of the great apes*. Cambridge, UK: Cambridge University Press.

Russon, A. E., & Begun, D. R. (2004). Evolutionary origins of great ape intelligence. In A. E. Russon & D. R. Begun (Eds.), *The evolution of thought: Evolutionary origins of great ape intelligence* (pp. 353–368). Cambridge, UK: Cambridge University Press.

Russon, A.E., van Schaik, C. P., Kuncoro, P., Ferisa, A., Handayani, P., & van Noordwijk, M. A. (2009). Innovation and intelligence in orangutans. In S. A. Wich, S. S. Utami Atmoko, T. Mitra Setia, & C. P. van Schaik (Eds.), *Orangutans: Geographic variation in behavioral ecology and conservation* (pp. 279–298). Oxford, UK: Oxford University Press.

Sanz, C. M., & Morgan, D. B. (2007). Chimpanzee tool technology in the Goualougo Triangle, Republic of Congo. *Journal of Human Evolution*, *52*, 420–433.

Savage-Rumbaugh, S., McDonald, K., Sevcik, R. A., Hopkins, W. D., & Rubert, E. (1986). Spontaneous symbol acquisition and communicative use by pygmy chimpanzees (*Pan paniscus)*. *Journal of Experimental Psychology: General, 115*, 211–235.

Schwartz, J. H. (1999). *Sudden origins*. New York, NY: Wiley.

Semaw, S., Renne, P., Harris, J. W. K., Feibel, C. S., Bernor, R. L., et al. (1997). 2.5-million-year-old stone tools from Gona, Ethiopia. *Nature, 385*, 333–336.

Shumaker, R. W., Palkovich, A. M., Beck, B. B., Guagnano, G. A., & Morowitz, H. (2001). Spontaneous use of magnitude discrimination and ordination by the orangutan (*Pongo pygmaeus*). *Journal of Comparative Psychology*, *115*, 385–391.

Sloman, S. (1996). The empirical case for two systems of reasoning. *Psychological Bulletin, 9*(1), 3−22.

Smith, W. M., Ward, T. B., & Finke, R. A. (1995). *The creative cognition approach*. Cambridge, MA: MIT Press.

Smolensky, P. (1988). On the proper treatment of connectionism. <u>Behavioral and Brain Sciences</u> *11*(1), 1-23.

Soffer, O. (1994). Ancestral lifeways in Eurasia – The middle and upper Paleolithic records. In M. Nitecki & D. Nitecki (Eds.), *Origins of anatomically modern humans*. New York, NY: Plenum Press.

Sperber, D. (1994). The modularity of thought and the epidemiology of representations. In L. A. Hirshfield & S. A. Gelman (Eds.), *Mapping the mind: Domain specificity in cognition and culture*. Cambridge, UK: Cambridge University Press.

Stanovich, K. E. (2005). *The robot's rebellion: Finding meaning in the age of Darwin*. Chicago, IL: University of Chicago Press.

Stanovich, K. E., & West, R. F. (2000). Individual differences in reasoning: Implications for the rationality debate? *Behavioral and Brain Sciences, 23*, 645–726.

Sternberg, R. J. (2001). Why schools should teach for wisdom: The balance theory of wisdom in educational settings. *Educational Psychologist*, *36*, 227–245.

Stringer, C., & Gamble, C. (1993). *In search of the Neanderthals.* London, UK: Thames and Hudson.

Suddendorf, T., & Whiten, A. (2002). Mental evolution and development: Evidence for secondary representation in children, great apes, and other animals. *Psychological Bulletin*, *127*, 629–650.

Swisher, C. C., Curtis, G. H., Jacob, T., Getty, A. G., Suprijo, A., et al. (1994). Age of the earliest known hominids in java, Indonesia. *Science, 263,* 118–121.

Thompson, R. K. R., & Oden, D. L. (2000). Categorical perception and conceptual judgments by nonhuman primates: The paleological monkey and the analogical ape. *Cognitive Science*, *24*, 363–396.

Tomasello, M., Kruger, A. C., & Ratner, H. H. (1993). Cultural learning. *Behavioral and Brain Sciences*, *16*, 495–552.

Tomasello, M. (1999). *The cultural origins of human cognition.* Cambridge, MA: Harvard University Press.

Tomasello, M., & Call, J. (1997). *Primate cognition*. New York, NY: Oxford University Press.

van Schaik, C. P., Ancrenaz, M., Borgen, G., Galdikas, B., Knott, C. D., Singleton, I., Suzuki, A., Utami, S. S., Merrill, M. (2003). Orangutan cultures and the evolution of material culture. *Science*, *299*, 102–105.

Vetsigian, K., Woese, C., & Goldenfeld, N. (2006). Collective evolution and the genetic code. *Proceedings of the New York Academy of Science USA, 103*, 10696–10701.

de Waal, F. B. M. (2001). *The ape and the sushi master: Cultural reflections by a primatologist*. New York, NY: Basic Books.

Walker, A. C. & Leakey, R. E. (1993). *The Nariokotome Homo erectus skeleton*. Cambridge, MA: Harvard University Press.

Werner, H. (1948). *Comparative psychology of mental development*. New York, NY: International Universities Press.
White, R. (1982). Rethinking the middle/upper Paleolithic transition. *Current Anthropology, 23*, 169–189.
White, R. (1989a). Production complexity and standardization in early Aurignacian bead and pendant manufacture: Evolutionary implications. In P. Mellars & C. Stringer (Eds.), *The human revolution: Behavioral and biological perspectives on the origins of modern humans* (pp. 366–90). Cambridge, UK: Cambridge University Press.
White, R. (1989b). Toward a contextual understanding of the earliest body ornaments. In E. Trinkhaus (Eds.), *The emergence of modern humans: Biocultural adaptations in the later Pleistocene.* Cambridge, UK: Cambridge University Press.
White, R. (1993). Technological and social dimensions of "Aurignacian-age" body ornaments across Europe. In H. Knecht, A. Pike-Tay, & R. White (Eds.), *Before Lascaux: The complex record of the early upper Paleolithc.* New York, NY: CRC Press.
White, T., Asfaw, B., Degusta, D., Gilbert, H., Richards, G. D., et al. (2003). Pleistocene Homo sapiens from middle awash, Ethiopia. *Nature, 423,* 742–747.
Whiten, A. (Ed.). (1991). *Natural theories of mind*. Oxford, UK: Basil Blackwell.
Whiten, A., & Byrne, R. (Eds.). (1997). *Machiavellian intelligence II: Extensions and evaluations*. Cambridge, UK: Cambridge University Press.
Whiten, A., Goodall, J., McGrew, W. C., Nishida, T., Reynolds, V., Sugiyama, Y., Tutin C. E. G., Wrangham, R. W., & Boesch, C. (1999). Culture in chimpanzees. *Nature*, *399,* 682–685.
Whiten, A., Schick, K., & Toth, N. (2009). The evolution and cultural transmission of percussive technology: Integrating evidence from palaeoanthropology and primatology. *Journal of Human Evolution*, *57*, 420–435.
Wilson, D. S., Near, D., & Miller, R. R. (1996). Machiavellianism: A synthesis of the evolutionary and psychological literatures. *Psychological Bulletin, 119*, 285–299.
Wundt, W. (1896). *Lectures on human and animal psychology*. New York, NY: Macmillan.
Wynn, T. (1998). Did Homo erectus speak? *Cambridge Archaeological Journal, 8,* 78–81.

---

## Footnotes

[1] The term *mimetic* is derived from "mime," which means "to act out."

[2] For those who think in neural net terms, contextual focus amounts to the capacity to spontaneously and subconsciously vary the shape of the activation function, flat for divergent thought and spiky for analytical.